\documentclass[conference]{IEEEtran}
\IEEEoverridecommandlockouts
\usepackage{graphicx}
\usepackage{cite}
\usepackage{amsmath}
\usepackage{amssymb}
\usepackage[square,sort&compress,comma]{natbib}
\usepackage{verbatim}
\usepackage{multirow}
\usepackage{diagbox}
\usepackage{booktabs}

\usepackage{algpseudocode}
\usepackage[ruled]{algorithm2e}
\usepackage{algorithm2e}

\usepackage{natbib}

\usepackage[
top    = 0.753in,
bottom = 0.98in,
left   = 0.673in,
right  = 0.673in]{geometry}

\def\ba{{\boldsymbol a}}
\def\bs{{\boldsymbol s}}
\def\by{{\boldsymbol y}}
\def\bw{{\boldsymbol w}}
\def\bI{{\boldsymbol I}}
\def\bH{{\boldsymbol H}}
\def\bF{{\boldsymbol F}}

\def\BibTeX{{\rm B\kern-.08em{\sc i\kern-.025em b}\kern-.08em
    T\kern-.1667em\lower.7ex\hbox{E}\kern-.125emX}}
\begin{document}

\title{DNN-Based Precoding in RIS-Aided mmWave MIMO Systems With Practical Phase Shift\\\vspace{-0.1in}
\thanks{This work was supported in part by the Academia Sinica (AS) under Grant 235g Postdoctoral Scholar Program, in part by the National Science and Technology Council (NSTC) of Taiwan under Grant 113-2218-E-110-008, 113-2218-E-110-009 and 113-2926-I-001-502-G, and in part by the U.S. National Science Foundation (NSF) under Grant CNS-2030215.}
}
\author{\IEEEauthorblockN{Po-Heng Chou$^{1}$, Ching-Wen Chen$^{2}$, Wan-Jen Huang$^{2}$, Walid Saad$^{3,4}$, Yu Tsao$^{1}$, and Ronald Y. Chang$^{1}$}
\IEEEauthorblockA{$^{1}$Research Center for Information Technology Innovation (CITI), Academia Sinica (AS), Taipei, 11529, Taiwan\\
$^{2}$Institute of Communication Engineering (ICE), National Sun Yat-sen University (NSYSU), Kaohsiung, 80424, Taiwan\\
$^{3}$Bradley Department of Electrical and Computer Engineering (ECE), Virginia Tech (VT),
Arlington, VA, 22203, USA\\
$^{4}$Artificial Intelligence \& Cyber Systems (AICS) Research Center, Lebanese American University (LAU), Lebanon\\
E-mails: d00942015@ntu.edu.tw, p19960716@gmail.com, wjhuang@faculty.nsysu.edu.tw,\\ walids@vt.edu, yu.tsao@citi.sinica.edu.tw, rchang@citi.sinica.edu.tw \vspace{-0.2in}}
}

\maketitle

\begin{abstract}
In this paper, the precoding design is investigated for maximizing the throughput of millimeter wave (mmWave) multiple-input multiple-output (MIMO) systems with obstructed direct communication paths.
In particular, a reconfigurable intelligent surface (RIS) is employed to enhance MIMO transmissions, considering mmWave characteristics related to line-of-sight (LoS) and multipath effects. 
The traditional exhaustive search (ES) for optimal codewords in the continuous phase shift is computationally intensive and time-consuming.
To reduce computational complexity, permuted discrete Fourier transform (DFT) vectors are used for finding codebook design, incorporating amplitude responses for practical or ideal RIS systems.
However, even if the discrete phase shift is adopted in the ES, it results in significant computation and is time-consuming.
Instead, the trained deep neural network (DNN) is developed to facilitate faster codeword selection.
Simulation results show that the DNN maintains sub-optimal spectral efficiency even as the distance between the end-user and the RIS has variations in the testing phase.
These results highlight the potential of DNN in advancing RIS-aided systems.

\end{abstract}


\IEEEpeerreviewmaketitle
\vspace{-0.1in}

\section{Introduction}

Millimeter wave (mmWave)~\cite{Haider2022} multiple-input multiple-output (MIMO)~\cite{Albreem2021} communication systems will be widely used in next-generation wireless networks due to their ability to transmit gigabit-per-second data rates by utilizing a large amount of bandwidth in mmWave frequencies (24--52 GHz). However, such an extremely high-frequency technology consumes a large amount of power and can suffer from blockage events easily in dense urban or indoor environments due to its small wavelength~\cite{Christina2022,Omar2023,Zhu2022}. To address this challenge, the use of a relay can help enhance the quality of service, reduce blockage probabilities, and expand the coverage of mmWave signals. Relays equipped with antenna arrays still require active elements and transmitters/receivers, so the implementation costs and power consumption of radio frequency (RF) chains are costly.

Recently, the use of a reconfigurable intelligent surface (RIS)~\cite{Wu2021,Zhu2022,Omar2023} has been proposed for economically improving the spectrum and energy efficiency. An RIS consists of several passive reflective elements and a controller. The details of RIS passive reflective elements material composition and implementation can be found in~\cite{Abeywickrama2020TWC} and~\cite{Abeywickrama2020ICC}. When the line-of-sight (LoS) link is obstructed by obstacles, the transmitted signal can bypass the obstacles through the RIS without suffering energy attenuation. Each passive reflective element can reflect the transmitted signal independently with a reconfigurable amplitude and phase shift via a software-defined controller, and no need to use any RF chains. 
The authors in~\cite{Cheonyong2022} asymptotically
analyze the spectral efficiency and show significant complexity
of the RIS-aided mmWave MIMO systems. 
There is a critical issue of how to joint design the phase shifts of the active beamforming and passive reflecting elements to obtain the array response that improves the spectral efficiency~\cite{Ruijin2023, Zhu2022,Wang2020, Abeywickrama2020TWC, Abeywickrama2020ICC}. However, the tightly coupled optimization
variables lead to the formulated problem being non-convex~\cite{Ruijin2023}. Due to the non-convex property, the traditional exhaustive search (ES) for finding optimal phase shifts is computationally intensive and time-consuming~\cite{Ruijin2023,Wang2023,Huang2023, Gong2022,Zhu2022}.
In particular, the global optimal solution to this problem in continuous phase shifts is a more arduous challenge.

Most of the previous studies~\cite{Wang2023,Huang2023,Omar2023,Wang2020,Huang2020} investigated the RIS optimization with independent amplitude and phase shift.
The authors in~\cite{Abeywickrama2020TWC} and~\cite{Abeywickrama2020ICC} propose the practical phase-dependent amplitude model to reflect the relationship between phase shift and amplitude of RIS elements.
The use of a practical phase-dependent amplitude of RIS leads to additional challenges and computational burdens in performing the optimization.
The main contribution of this paper is joint design active and passive precoding is investigated for maximizing the throughput of RIS-Aided mmWave MIMO systems with a practical phase-dependent amplitude model~\cite{Abeywickrama2020TWC} and~\cite{Abeywickrama2020ICC}. 
In addition, we generate a discrete codebook based on the Kronecker product of beam-steering vectors corresponding to azimuth angle and elevation angle for lower computational complexity. 
Specifically, each discrete codeword corresponds to a set of phase shifts performed at RIS, and our goal is to choose the best codeword from the codebook in terms of maximal achievable rate. 
We use a deep neural network (DNN) to search for the near-optimal discrete codeword, which is more efficient than ES~\cite{Ozpoyraz2022}. Given the reflective model of the RIS, the proposed DNN can capture the variation in the amplitude response and generate a near-optimal discrete codeword corresponding to the current channel status. Simulation results show that the proposed DNN can achieve near-optimal spectral efficiency ($99.68\%$). Meanwhile, this approach simply takes $3.5\%$ of computational time compared with the ES.
\vspace{-0.1in}
\section{System and Channel Models}
\begin{figure}[tb]
\centering
{\includegraphics[width=0.45\textwidth]{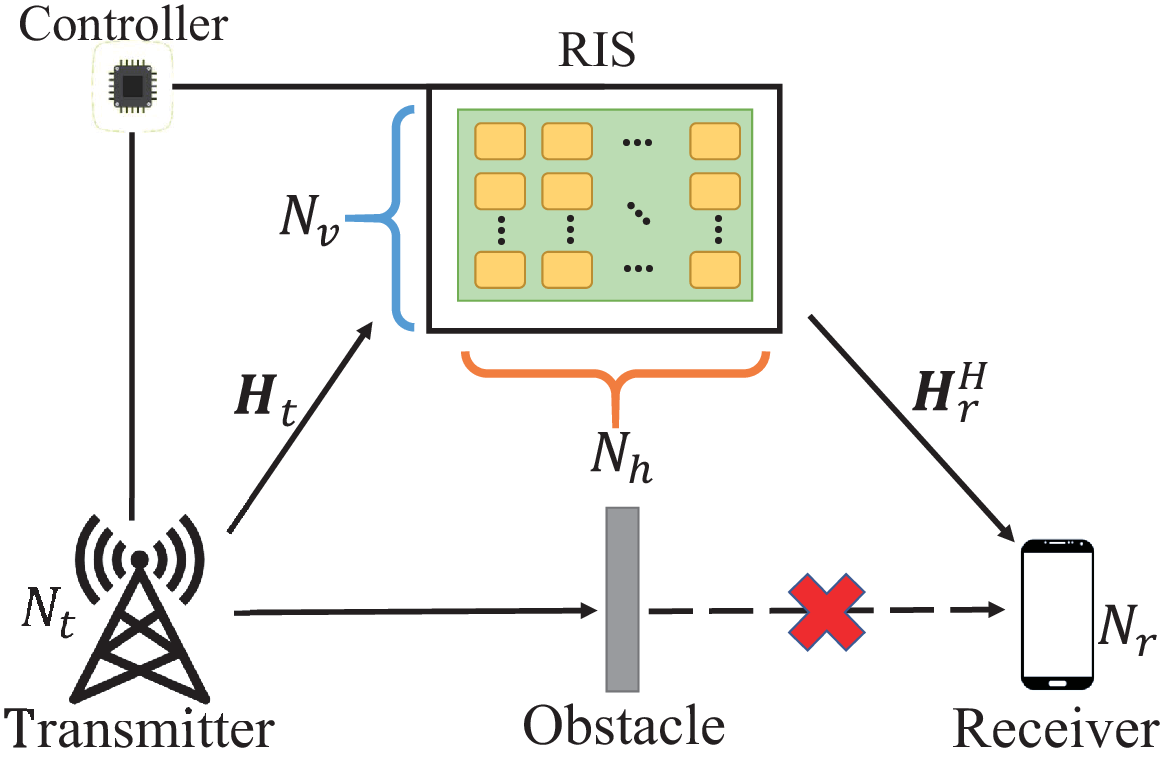}
\caption{The RIS-aided mmWave MIMO system.}\label{System_Model}}
\vspace{-0.2in}
\end{figure}

Consider a MIMO system consisting of a single transmitter and receiver equipped with
$N_t$ and $N_r$ antennas, respectively, as illustrated in Fig.~\ref{System_Model}. Assume that the direct path from the transmitter to the receiver is obstructed, and the transmission is aided by a RIS system with $N = N_{h} \times N_{v}$ passive reflecting elements, where $N_h$ and $N_v$ are respectively numbers of horizontal columns and vertical rows in the RIS array.
\vspace{-0.03in}
Let $\bs$ be an $N_s\times 1$ vector of data streams with covariance matrix $E[\bs\bs^H]=\frac{P}{N_{s}}\bI_{N_{s}}$, where $P$ is the total transmission power. The transmitted signal arrives at the receiver after being reflected by the RIS. The equivalent baseband signal at the receiver is given by
\begin{equation}\label{eq:rx_sig}
\by = \bH_{r}^{H}\boldsymbol{\Psi}\bH_{t}\bF\bs + \bw,
\end{equation}
where $\boldsymbol{F} \in \mathbb{C}^{N_{t} \times N_{s}}$ is a precoding matrix, $\boldsymbol{\Psi} \in \mathbb{C}^{N \times N}$ is a diagonal matrix of the RIS response,
$\boldsymbol{H}_{t} \in \mathbb{C}^{N \times N_{t}}$ and $\boldsymbol{H}_{r}^{H} \in \mathbb{C}^{N_{r} \times N}$ are the channel matrices from the transmitter to the RIS and from the RIS to the receiver, respectively, and $\boldsymbol{w}\thicksim {\cal CN}(0, \sigma^{2}\bI_{N_{s}})$ is the additive white Gaussian noise (AWGN) vector.

We consider the transmitter and receiver to be both equipped with uniform linear array (ULA), and the transmission is over the mmWave spectrum and the LoS exists between the transmitter and the RIS or between the RIS and the receiver. Hence, the channel matrices
can be given by:
\begin{align} 
\bH_{t} &= \sqrt{L_{t}} \left(\sqrt{\frac{K_{t}}{K_{t}+1}}\overline{\bH}_{t}+\sqrt{\frac{1}{K_{t}+1}}\widetilde{\bH}_{t}\right),\label{eq:Ht}\\
\bH_{r}^{H} &= \sqrt{L_{r}}\left(\sqrt{\frac{K_{r}}{K_{r}+1}}\overline{\bH}^{H}_{r}+\sqrt{\frac{1}{K_{r}+1}}\widetilde{\bH}_{r}^{H}\right),\label{eq:Hr}
\end{align}
where $L_t$ and $L_r$ are respectively the path loss of the transmitter-RIS link and the RIS-receiver link, and $K_t$ and $K_r$ are Rician factors of the transmitter-RIS link and the RIS-receiver link.
The path loss factors are given by
\begin{align*}
L_{t}({\rm dB}) &= L_{0}({\rm dB}) - 10\xi_{t}\log_{10}\left(\frac{d_{t}}{D_{0}}\right) + G_{\rm RIS}({\rm dB}),\\
L_{r}({\rm dB}) &= L_{0}({\rm dB}) - 10\xi_{r}\log_{10}\left(\frac{d_{r}}{D_{0}}\right) + G_{\rm RIS}({\rm dB}),
\end{align*}
where $d_t$ and $d_r$ are the distances between the transmitter and the RIS and between the receiver and the RIS, respectively, $\xi_t$ and $\xi_r$ are the path loss exponents of the transmitter-RIS link and the RIS-receiver link, $L_{0}$ is the path loss with the reference distance $D_0$, and
$G_{\rm RIS}$ is gain of the RIS. 
In (\ref{eq:Ht}), $\overline{\boldsymbol{H}}_{t}$ and $\widetilde{\boldsymbol{H}}_{t}$ are respectively the LoS component and non-line-of-sight (NLoS) component of the channel matrix $\boldsymbol{H}_{t}$,  given by
\begin{align}
\overline{\bH}_{t} &= [\ba^{H}_{N_{h}}(\phi_{h,0})\otimes \ba^{H}_{N_{v}}(\phi_{v,0})]\ba_{N_{t}}(\theta_{t,0}),\\
\widetilde{\bH}_{t} &= \sum_{\ell =1}^{L} z_\ell[\ba^{H}_{N_{h}}(\phi_{h,\ell})\otimes \ba^{H}_{N_{v}}(\phi_{v,\ell})]\ba_{N_{t}}(\theta_{t,\ell}),
\end{align}
where $\otimes$ is the Kronecker product of two vectors, $\theta_{t,\ell}=2\pi\frac{d}{\lambda}\sin(\theta_{\rm AoD,\ell})$, 
$\phi_{h,\ell}=2\pi\frac{d}{\lambda}\sin(\phi_{\rm aAoA,\ell})$, 
and $\phi_{v,\ell}=2\pi\frac{d}{\lambda}\sin(\phi_{\rm eAoA,\ell})$ with
$\theta_{\rm AoD,\ell}$, $\phi_{\rm aAoA,\ell}$ and $\phi_{\rm eAoA,\ell} \in U[0, \pi]$ being the angle of departure (AoD) of the $\ell$-th path at the transmitter,  the azimuth angle of arrival (AoA), and the elevation AoA of the $\ell$-th path at the RIS, respectively. Note that path $\ell=0$ refers to the LoS path. In addition, the parameter $z_{\ell}\sim {\cal CN}(0, 1/L)$ is the complex fading gain of the $\ell$-th NLoS path, $\lambda$ is the signal wavelength and $d$ is the antenna spacing, which is set as $d = \lambda/2$. The beam-steering vector $\boldsymbol{a}_{N}(\theta)\in \mathbb{C}^{N\times 1}$ is defined by $\boldsymbol{a}_{N}(\theta)
= \left[1, e^{-j\theta}, \ldots , e^{-j(N-1)\theta}\right].$

Similarly, the LoS and NLoS components of the channel between the RIS and the receiver can be written by
\begin{align}
\overline{\boldsymbol{H}}_{r}^{H} &= \boldsymbol{a}^{H}_{N_{r}}(\theta_{r,0})[\boldsymbol{a}_{N_{v}}(\varphi_{v,0})\otimes \boldsymbol{a}_{N_{h}}(\varphi_{h,0})],\\
\widetilde{\boldsymbol{H}}_{r}^{H} &= \sum_{\ell = 1}^{L} z_{\ell}\boldsymbol{a}^{H}_{N_{r}}(\theta_{r,\ell})[\boldsymbol{a}_{N_{v}}(\varphi_{v,\ell})\otimes \boldsymbol{a}_{N_{h}}(\varphi_{h,\ell})],
\end{align}
where $\theta_{r,\ell}=2\pi\frac{d}{\lambda}\sin(\theta_{\rm AoA,\ell})$, 
$\varphi_{v,\ell}=2\pi\frac{d}{\lambda}\sin(\varphi_{\rm eAoD,\ell})$,
and $\varphi_{h,\ell}=2\pi\frac{d}{\lambda}\cos(\varphi_{\rm eAoD,\ell})\sin(\varphi_{\rm aAoD,\ell})$ with
$\theta_{\rm AoA,\ell}$, $\varphi_{\rm \rm aAoD,\ell}$ and $\varphi_{\rm eAoD,\ell}$ being the AoA of the $\ell$-th path at the receiver,  the azimuth AoD and the elevation AoD of $\ell$-th path at the RIS, respectively.

The reflective elements of the RIS can shift the phase of the incident signal to exploit the array gain. The RIS response matrix is given by 
$\boldsymbol{\Psi} = \textrm{diag} (\boldsymbol{\Phi}) = \textrm{diag}(\beta_{1}e^{j\psi_{1}}, \beta_{2}e^{j\psi_{2}}, \ldots , \beta_{N}e^{j\psi_{N}})$, where $\textrm{diag} (\cdot)$ is vector-to-diagonal matrix operator, $\boldsymbol{\Phi}$ is the RIS response vector, $\beta_{n} \in (0, 1]$ and $\psi_n \in [- \pi, \pi]$ represent the amplitude response and the phase shift of the $n$-th RIS reflective element, and $n = 1, \ldots , N$. Ideally, the amplitude response is uniform and $\beta_n = 1, \forall n$. Practically, the amplitude response may vary with the phase shifting due to the non-zero impedance of the reflective circuit. To take this practical issue into consideration, we adopt the approximated amplitude response \cite{Abeywickrama2020TWC, Abeywickrama2020ICC} given by
\begin{align}
\beta_n = (1-\beta_{\min})\left(\frac{\sin(\psi_{n}-\psi_0)+1}{2}\right)^{\alpha}+\beta_{\min},
\end{align}
for $\psi \in (-\pi, \pi]$, where $\beta_{\min}$ is the minimum amplitude, $\psi_0$ is the difference between $0^{\circ}$ (the phase with minimum amplitude) and $-\pi$, and $\alpha$ controls the steepness of the function curve. When $\alpha = 0$, it reduces to the ideal case of reflective amplitude.

\section{Problem Formulation and Codebook Design}

\subsection{Problem Formulation}
Our goal is to design the transmit precoder as well as phase-shifting parameters of the RIS to maximize the spectral efficiency. 
We define $\boldsymbol{H}_{\rm eff} = \boldsymbol{H}_{r}^{H}\boldsymbol{\Psi}\boldsymbol{H}_{t}$ as the equivalent channel between transmitter and receiver. To obtain the maximal spatial multiplexing gain, we assume that $N_{s} = \textrm{rank}(H_{\rm eff})$. To maximize the achievable rate, we formulate the following optimization problem:
\begin{subequations}
\begin{align}
\max_{\boldsymbol{F}, \boldsymbol{\Psi}} \;\;\log_{2}&\left|\boldsymbol{I}_{N_{s}}+\frac{P}{\sigma^{2}N_{s}}\boldsymbol{H}_{\rm eff}^{H}\boldsymbol{F}\boldsymbol{F}^{H}\boldsymbol{H}_{\rm eff}\right|, \label{optimization_problem}\\
\textrm{s.t}\;\;\|{\boldsymbol F} \|^{2}_{F} &\leq N_{s}\\
[\boldsymbol{\Psi}]_{i,j} &=
\left\{
\begin{array}{ll}
\beta(\psi_{i})e^{-j\psi_{i}},& i =j\\
0,&  i\neq j.
\end{array}\right.
\end{align}
\label{optimization_problem_all}
\end{subequations}
\hspace{-0.063in}where $\left|\cdot\right|$ is the determinant operator, the first constraint limits the maximum transmission power, and the second constraint follows the reflective response of the RIS.  
The optimization problem is non-convex because of the constraint on the RIS reflective matrix. 
To solve this intractable problem, we may optimize the precoder and phase-shifting parameters by alternating optimization (AO). However, AO introduces significant computational complexity and lengthy processing times because of its iterative process, and it may yield sub-optimal solutions~\cite{Ruijin2023}. Hence, we proposed an alternative deep learning (DL)-based approach, which only requires a single execution. 
Given the reflecting matrix $\boldsymbol{\Psi}$, the optimal precoding matrix can be obtained by performing the singular value decomposition (SVD) on the effective channel matrix $\boldsymbol{H}_{\rm eff}$. Specifically, we define $\boldsymbol{H}_{\rm eff}$ the SVD of the effective channel matrix, where $\boldsymbol{U}\in \mathbb{C}^{N_{r} \times N_{s}}$ and $\boldsymbol{V}\in \mathbb{C}^{N_{t} \times N_{s}}$ are respectively the left and right singular matrices and  
$\boldsymbol{\Lambda} = \textrm{diag}(\tau_{1}, \tau_{2}, \ldots , \tau_{N_{s}})$ comprises of singular values on the diagonal.
The optimal precoding matrix is given by $\boldsymbol{F}_{\rm opt} = \boldsymbol{V}$,
and the corresponding  achievable  rate can be simplified by
$R= \sum_{c = 1}^{N_{s}} \log_{2}\left( 1 + \frac{P}{\sigma^{2}N_{s}}\tau_{c}\right).$
The achievable rate is now related to singular values of the effective channel matrix, which is linear in the RIS response matrix. 
Therefore, we can reformulate problem~(\ref{optimization_problem_all}) as
\begin{subequations}
\begin{align}
\max_{\boldsymbol{\Psi}} \;\;\sum_{c = 1}^{N_{s}}& \log_{2}\left( 1 + \frac{P}{\sigma^{2}N_{s}}\tau_{c}\right). \label{optimization_problem_2}\\
\textrm{s.t}\;\;
[\boldsymbol{\Psi}]_{i,j} &=
\left\{
\begin{array}{ll}
\beta(\psi_{i})e^{-j\psi_{i}},& i=j \\
0,& i \neq j.
\end{array}\right.
\end{align}
\label{optimization_problem_2_all}
\end{subequations}
However, the simplified optimization problem of single parameter $\boldsymbol{\Psi}$ in~(\ref{optimization_problem_2_all}) is still the non-convex that is difficult to solve. 

\subsection{Codebook Design}

To facilitate the precoding design, we choose an optimal set of phase-shifting parameters ${\rm diag}(\boldsymbol{\Psi}^*)$ within a predefined codebook. Since the channel matrices are composed of beam-steering vectors in our framework, we utilize the beam-steering vectors corresponding to quantized angles to construct the codebook.
Denote a steering vector w.r.t. angle $\theta$ as $\boldsymbol{g}(N, \theta) = [1, e^{-j\theta}, \ldots , e^{-j\theta (N-1)}]$. We quantize the azimuth and elevation angle as $\{\theta_{h,1}=0, \theta_{h,2}=Q_{h}, \ldots, \theta_{h,N_{h}}=(N_{h}-1)Q_{h}\}$ and $\{\theta_{v,1}=0, \theta_{v,2}=Q_{v}, \ldots, \theta_{v,N_{v}}=(N_{v}-1)Q_{v}\}$, respectively,
where $Q_{h}=\frac{2\pi}{N_{h}}$ and $Q_{v}=\frac{2\pi}{N_{v}}$ are respectively the quantization spacing of the azimuth and elevation angles. For the case with ideal RIS reflective response, the codebook $\mathcal{P}_{\rm ideal}$ can be constructed from the following vectors
$
\boldsymbol{a}_{N_{h}}(\theta_{h,i}) \otimes \boldsymbol{a}_{N_{v}}(\theta_{v,j})\in \mathcal{P}_{\rm ideal},
$
for $i=1,2,\ldots, N_h$, $j=1,2,\ldots, N_v$.
For practical consideration, the codewords shall take the amplitude response into consideration. Let ${\boldsymbol p}\in\mathcal{P}_{\rm ideal}$ be a codeword for the ideal case, which is denoted as
${\boldsymbol p}\triangleq [e^{j\theta_{1}}, e^{j\theta_{2}}, \ldots, e^{j\theta_{N}}]^{T}$. Then the corresponding amplitude response is given by $\Upsilon({\boldsymbol p}) = [\beta_{1}, \beta_{2}, \ldots, \beta_{N}]^{T}$. The codebook that considers the practical amplitude response can be written as
$\mathcal{P}_{\rm prac} = \{\Upsilon(\boldsymbol{p})\odot \boldsymbol{p} : \boldsymbol{p} \in \mathcal{P}_{\rm ideal}\},
$
where $\odot$ is the element-wise Hadamard product.

\section{Deep Neural Network (DNN)}
In this paper, we adopt a DNN with supervised learning, to determine the optimal RIS response.
We denote $\tilde{\boldsymbol{h}}_{\rm eff}={\rm vec}(\boldsymbol{H}_{\rm eff})\in\mathbb{C}^{N_tN_r\times 1}$ as a vectorization of $\boldsymbol{H}_{\rm eff}$, which can be given by
\begin{align}
\tilde{\boldsymbol{h}}_{\rm eff}^T = \boldsymbol{\Phi}^{T}\!\Big[\boldsymbol{h}_{r,1}^{H}\!\odot\! \boldsymbol{h}_{t,1}, \ldots , \boldsymbol{h}_{r,1}^{H}\!\odot\! \boldsymbol{h}_{t,N_{t}}, \boldsymbol{h}_{r,2}^{H}\!\odot\! \boldsymbol{h}_{t,1}, \ldots &\nonumber\\
\ldots,\boldsymbol{h}_{r,2}^{H}\!\odot\! \boldsymbol{h}_{t,N_t},\ldots\ldots ,\boldsymbol{h}_{r,N_{r}}^{H}\!\odot\! \boldsymbol{h}_{t,N_{t}}\Big],&
\end{align}
where $\boldsymbol{h}_{r,n}$ is the $n$-th row of channel matrix ${\boldsymbol H}_r$, and $\boldsymbol{h}_{t,m}$ is the $m$-th column of the channel matrix ${\boldsymbol H}_t$, 
for $n = 1, \ldots , N_{r}$ and $m = 1, \ldots , N_{t}$. Next, we define $\tilde{\boldsymbol{h}}_{\rm eff}^T\triangleq \boldsymbol{\Phi}^{T} {\mathcal H}$, where ${\mathcal H}$ comprises composite channel coefficients
from the $m$-th transmit antenna to the $n$-th receive antenna via each of the RIS element. In the DNN framework, we stack the real part and imaginary part of ${\mathcal H}$ as the input data vector,
which is pre-processed as follows: \textbf{Step 1.} Performing the Hadamard products of the channel state information (CSI) between the transmitter/receiver and RIS.
\textbf{Step 2.} Arranging the results of the Hadamard product operation into a complex vector.
\textbf{Step 3.} Separating and rearranging the real part and imaginary part of the complex vector as the input vector for the DNN model.
\textbf{Step 4.} Computing~(\ref{optimization_problem_2}) by using ES. Finding the optimum codeword vector that achieves the maximum spectrum efficiency corresponding to CSI in the codebook as the label.

\begin{figure}[t]
\centering
{\includegraphics[width=0.5\textwidth]{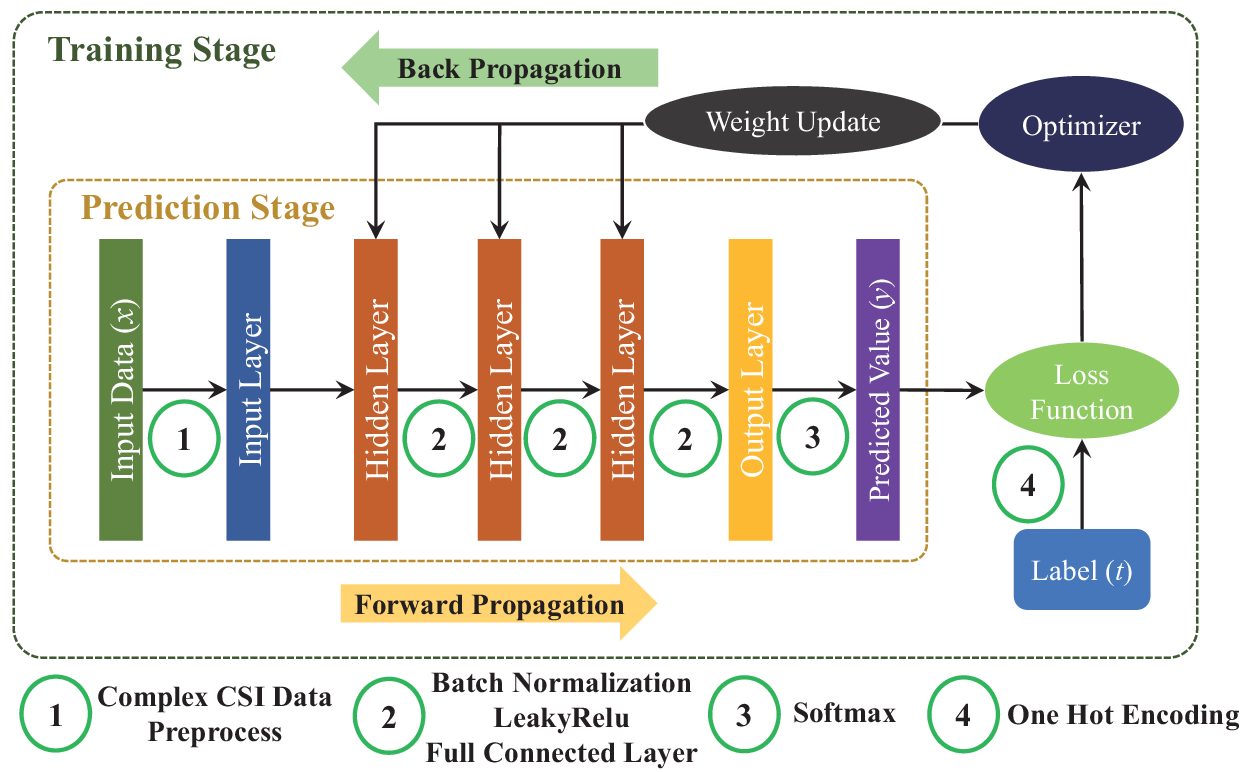}
\caption{DNN model training flow chart.}\label{DNN}}
\vspace{-0.31in}
\end{figure}

Our DNN~\cite{Ozpoyraz2022} comprises an input layer, three hidden layers, and an output layer.
The training phase is accomplished by two procedures, including forward propagation (prediction stage) and back-propagation (training stage), as shown in Fig.~\ref{DNN}.
In the stage of forward propagation, the tasks performed in each layer are described as follows:
\textbf{Input layer:} In general, neural networks only consider real values. Thus, we separate the complex value CSI data (Hadamard products of the CSI between the transmitter/receiver and RIS) into real and imaginary parts and arrange the imaginary part behind the real part. Then, the length of an input data column vector is $2N_{t} N_{r} N_{h} N_{v} \times 1$. 
\textbf{Hidden layers:} The activation value of each neuron is the sum of the activation value of the previous layer multiplied by the corresponding weight and added the corresponding bias.
The adjacent hidden layers are fully connected layers.
In the next step, batch normalization is used to normalize each mini-batch with standard normal distribution ${\cal N}(0, 1)$. 
The activation function (i.e., LeakyReLu) is then applied to convert linear input into nonlinear output.
\textbf{Output layer:} The main task of this study is to use the input (CSI data) to determine the label (the optimum codeword vector address). The output values of candidate codeword vectors corresponding to the input data are presented between 0 and 1 with a summation to be 1. By selecting the highest probability value through the activation function (i.e., Softmax), it is predicted as the optimum codeword vector.

The back-propagation is accomplished by the following procedures: \textbf{Calculation of the loss function:} When the model obtains the predicted value, it computes the error (loss) between the predicted value and label. In the classification problem, we use cross-entropy as the loss function. 
To calculate cross-entropy, we use one-hot encoding~\cite{one-hot2021}, a common method to convert categorical labels of codeword into binary value vectors. The loss function can be expressed as $L = -\frac{1}{B}\sum_{b=1}^{B}\sum_{n=1}^{N}t_{b,n}\ln (y_{b,n})$,
where $B$ is the batch size, $N = N_h \times N_v$ is the codebook size, the same as the number of output nodes, $y_{b,u} \in [0,1]$ is the output (predicted value), and $t_{b,u}$ is the label (codeword vector). \textbf{Optimizer:} The weights and biases are adjusted by the gradient descent to reduce the loss value. To adjust the learning rate based on the gradient and avoid finding local optima, we use Adam as the optimizer.
\vspace{-0.05in}
\section{Simulation Results}
For our simulations, the numbers of transmitted antennas and received antennas are  $N_{t}=10$ and  $N_{r}=1$ or $2$, respectively. The transmission power  $P$ is $20$ dBm, and the noise variance $\sigma^{2}$ is $-80$ dBm. The number of RIS elements  is set as $N\in\{45, 65, 85, 105\}$, where $N_v=5$ and  $N_{h}\in\{9, 13, 17, 21\}$, correspondingly. The RIS gain $G_{\rm RIS}$ is 5 dB. The 
parameters of the practical reflective model are set as $\beta_{\min} = 0.2$, $\alpha = 1.6$, and $\psi =0.43\pi$.
For the propagation channel, the number of NLoS paths is set as $L=2$ with the Rician factor $K_{t} = K_{r}=10$. The path loss exponent $\xi_{t}$ and the distance $d_{t}$ between the transmitter and RIS are 2 and 10 m, respectively. The path loss exponent $\xi_{r}$ and the distance $d_{r}$ between the RIS and receiver are 2.8 and 30 m, respectively. The DNN parameters and hyperparameter settings are shown in Table~\ref{DNN_Parameters}.
To reduce the feedback information, we create a codebook for RIS design by merging and rearranging the discrete Fourier transform (DFT) vectors.
We compare the proposed DNN with two baselines that adopt the same DFT codebook: \textbf{(1) ES (upper bound):} It assesses all possible codewords from the DFT codebook to identify the optimal codeword that maximizes spectral efficiency at the receiver. \textbf{(2) Random selection (lower bound):} It randomly chooses a codeword from the DFT codebook, and each selected codeword has an equal probability of being chosen. We consider the same ideal ($\mathcal{P}_{\rm ideal}$) and practical ($\mathcal{P}_{\rm prac}$) DFT codebooks in the above schemes.


\begin{table}[t]
\centering \caption{DNN parameters and hyperparameters setting.}
\scriptsize
\begin{tabular}{|l|c|l|c|}
\hline
Input Layer                  &720/1040/1360/1680 &Training Set &$3 \times 10^{7}$ \\
\hline
Hidden Layer 1                 &360/520/680/840 &Validation Set               &$10\%$ of training Set\\
\hline
Hidden Layer 2                &180/260/340/420 &Testing Set                  &$10^{5}$\\
\hline
Hidden Layer 3               &90/130/170/210 &Batch Size                    &$2000$\\
\hline
Output Layer                 &45/65/85/105 &Learning Rate                &$5 \times 10^{-4}$\\
\hline
Optimizer                    &Adam &Early Stopping               &2\\
\hline
\end{tabular}
\label{DNN_Parameters}
\vspace{-0.1in}
\end{table}

\begin{figure}[t]
\centering
{\includegraphics[width=0.4\textwidth]{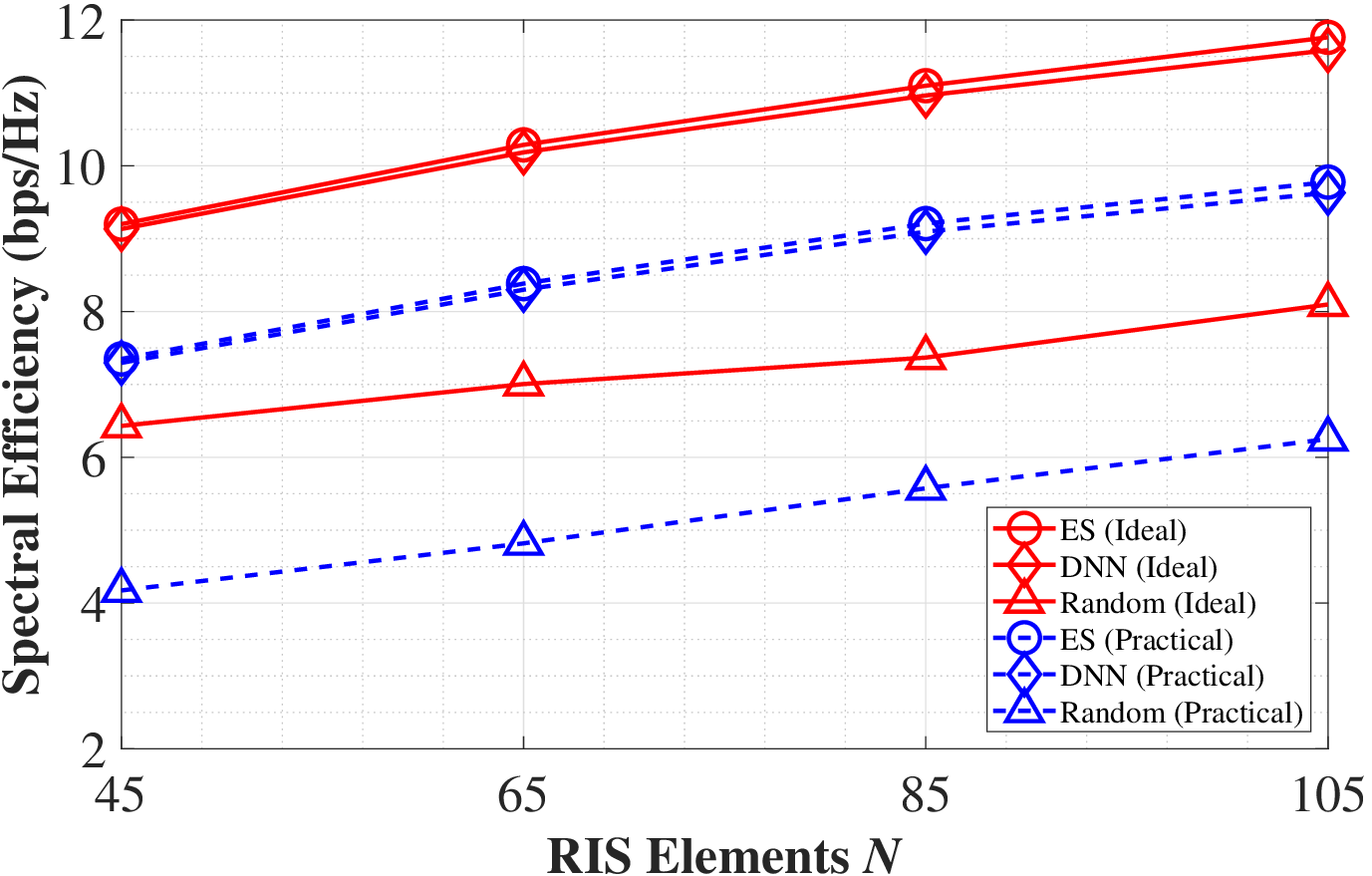}
\caption{The spectral efficiency in the different RIS element numbers under the mmWave channel assumption ($K_t = K_r = 10$, $N_{t} = 10$, $N_{r} = 2$, and $L = 2$).}\label{RIS}}
\vspace{-0.2in}
\end{figure}

Fig.~\ref{RIS} shows the spectral efficiency in terms of RIS element numbers under the mmWave channel assumption.
The performance of the proposed DNN can achieve between $98.8$--$99.6\%$ of ES and outperform the random selection by about 4 bps/Hz. With different numbers of RIS elements ($N = 45, 65, 85, 105$), the DNN outperforms random selection under ideal ($42\%$, $45.4\%$, $48.8\%$, $43\%$) and practical ($74.9\%$, $72.2\%$, $63.1\%$, $54\%$) conditions.

\begin{figure}[t]
\centering
{\includegraphics[width=0.4\textwidth]{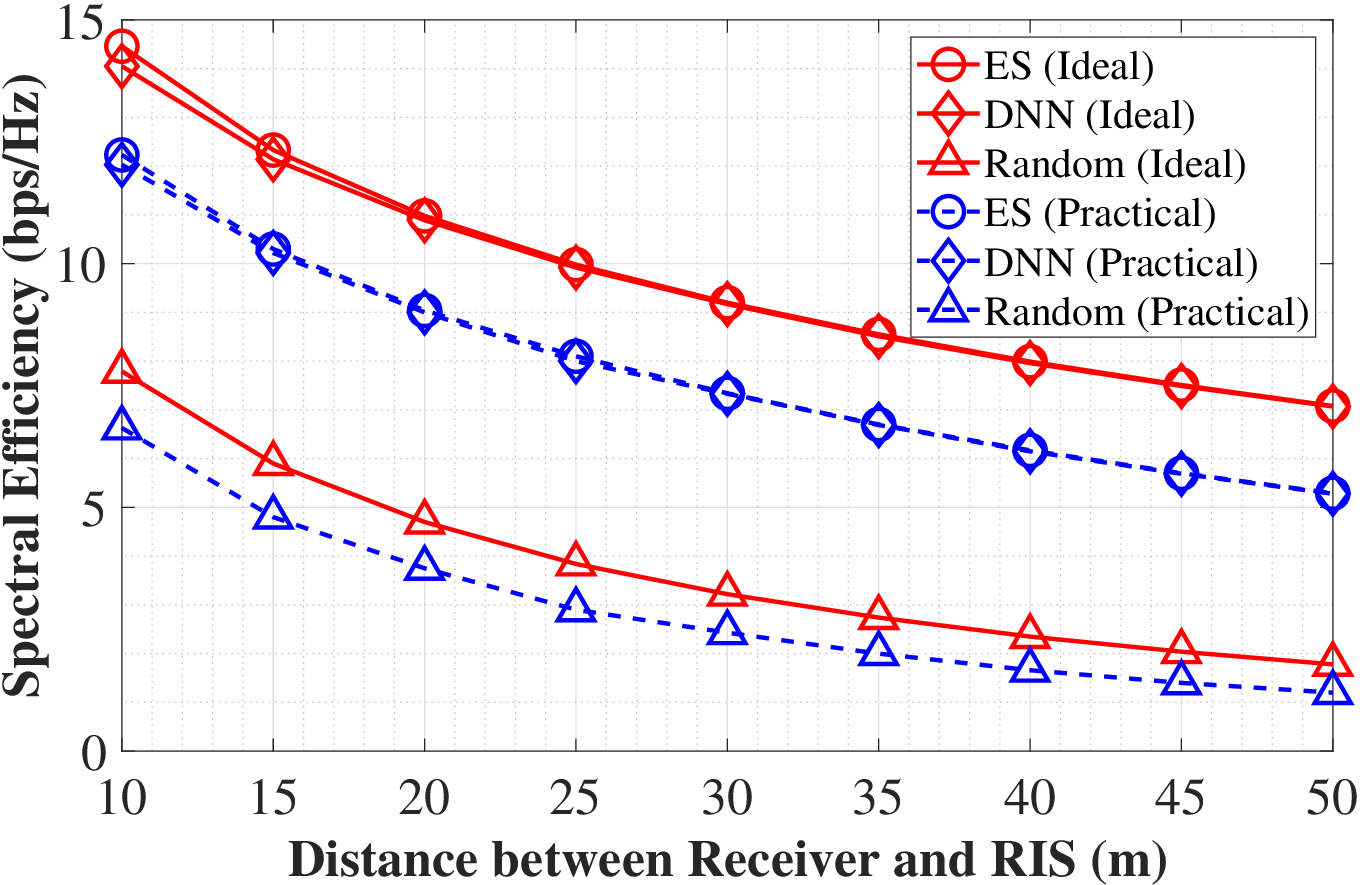}
\caption{The spectral efficiency at different distances between the receiver and RIS ($K_t = K_r = 10$, $N_{t} = 10$, $N_{r} = 2$, $N = 45$ and $L = 2$).}\label{distance}}	
\vspace{-0.1in}
\end{figure}

Fig.~\ref{distance} plots the spectral efficiency in terms of the distance between the receiver and RIS in different cases.
The spectral efficiency decreases with increasing distance. The performance of the proposed DNN can achieve $97.2$--$99.79\%$ of ES and outperform random selection by about 6 bps/Hz.
Across various receiver distances ($d_r = 10, 20, 30, 40, 50$), the DNN outperforms random selection by substantial margins, with gains of $80.4\%$, $131.7\%$, $184.8\%$, $238.7\%$, $297.2\%$ under ideal conditions and $81.4\%$, $140\%$, $201.6\%$, $270.5\%$, $340\%$ in practical conditions, respectively.
This shows that when the codeword vector is not specifically designed, there is a substantial degradation in spectral efficiency. In addition, the spectral efficiency difference between the ideal and practical codebooks is 2 bps/Hz under all of the schemes as shown in Figs.~\ref{RIS} and~\ref{distance}.
Under practical considerations, the performance degrades due to amplitude reduction when the phase-dependent amplitude of the RIS deviates from $-\pi$ or $\pi$. Our approach involves training the DNN model using CSI from a single scenario ($d_t$ = 10 m and $d_r$ = 30 m). Then, we apply the proposed DNN model to other distance cases, yielding results that are sub-optimal and close to the optimal (ES) solution. We can notice the variations in performance between the DNN and random selection are much more pronounced compared to ideal conditions. This highlights that the proposed DNN excels in handling phase-dependent RIS amplitude under practical considerations. This success can be attributed to the DNN model's ability to effectively learn and capture the regularity of mmWave channel characteristics.

Table~\ref{mmWave_comparison} considers the same parameter settings as in Fig.~\ref{distance}. As observed, the proposed DNN significantly reduces computation time by two orders of magnitude while maintaining a performance level of $98.5\%$ when compared to the ES. This suggests the effectiveness of the proposed DNN in efficiently finding near-optimal solutions for the challenging problem of maximizing spectral efficiency in RIS-aided mmWave MIMO systems.

\begin{table}[t]
   \caption{Comparison of the computation time (sec.) and spectral efficiency under the mmWave channel assumption.} 
   \label{tab:example}
   \scriptsize
   \centering
   \begin{tabular}{|l|c|c|c|c|}
   \hline
   \textbf{Number of RIS elements $N$} &$45$  &$65$ &$85$ &$105$\\ 
   \hline
   Computation time of ES &0.00191 &0.00207  &0.0023 &0.0026\\
   \hline
   Computation time of DNN &0.000079 &0.000096 &0.00009  &0.000091\\
   \hline
   Spectral efficiency (DNN/ES) &$99.68\%$	&$99.31\%$	&$99.16\%$	&$98.8\%$\\
   \hline
   \end{tabular}
   \label{mmWave_comparison}
   \vspace{-0.15in}
\end{table}


\vspace{-0.1in}
\section{Conclusion}
In this paper, we have considered a MIMO system aided with RIS in the mmWave band to improve the transmission quality and reduce the blockage probability. 
The proposed RIS precoding design includes phase quantization in horizontal and vertical directions, combining corresponding codewords in the codebook. 
The correlation between phase and amplitude is considered in actual RIS, as supervised knowledge. 
Thus, the proposed DNN can approximate ES to significantly reduce computational complexity and time.

\ifCLASSOPTIONcaptionsoff
  \newpage
\fi

\end{document}